%% file: main.tex
\documentclass[conference]{IEEEtran}
\IEEEoverridecommandlockouts
\usepackage{cite}
\usepackage{amsmath,amssymb,amsfonts}
\usepackage{algorithmic}
\usepackage[shortlabels]{enumitem}
\usepackage{graphicx}
\usepackage{textcomp}
\usepackage{xcolor}
\usepackage{listings}
\usepackage{float} 
\usepackage{todonotes}
\usepackage{pifont}
\def\BibTeX{{\rm B\kern-.05em{\sc i\kern-.025em b}\kern-.08em
    T\kern-.1667em\lower.7ex\hbox{E}\kern-.125emX}}
\usepackage{multirow}
\usepackage{hhline}

\usepackage{microtype}
\usepackage{graphicx}
\usepackage{subcaption}
\usepackage{booktabs} 
\usepackage{todonotes}
\usepackage{pifont}
\usepackage{makecell}

\usepackage{amssymb}
\usepackage{pifont}

\usepackage{hyperref}

\usepackage{xcolor}

\usepackage{afterpage}

\newcommand{\good}[0]{\ding{51}}
\newcommand{\bad}[0]{\ding{55}}

\newcommand{\nota}[0]{$-$}
\newcommand{\papername}[1]{#1}
\newcommand{\llama}[0]{{Llama-3}~}

\usepackage{arydshln} 

\usepackage{xspace}

\usepackage{amsthm}
\newtheoremstyle{boldheadonly}
  {\topsep}    
  {\topsep}    
  {\itshape} 
  {}           
  {\bfseries}  
  {.}          
  { }          
  {}           

\theoremstyle{boldheadonly}

\definecolor{opcolor}{RGB}{255,230,150}     
\definecolor{tensorcolor}{RGB}{180,220,255} 

\newcommand{\low}[0]{{\Large $\bullet\circ\circ$}}
\newcommand{\medium}[0]{{\Large $\bullet\bullet\circ$}}
\newcommand{\high}[0]{{\Large $\bullet\bullet\bullet$}}
\begin{document}



\title{Evaluating Cross-Architecture Performance Modeling of Distributed ML Workloads Using StableHLO}
\author{
\IEEEauthorblockN{
Jonas Svedas*, Nathan Laubeuf$^\dagger$, Ryan Harvey*, Arjun Singh*, Changhai Man$^\ddagger$, \\
Abubakr Nada$^\dagger$, Tushar Krishna$^\ddagger$, James Myers*, Debjyoti Bhattacharjee$^\dagger$
}
\IEEEauthorblockA{\footnotesize
\textit{*imec, 20 Station Road, Cambridge CB1 2JD, UK}~
\textit{$^\dagger$imec, Kapeldreef 75, 3001 Leuven, Belgium}\\
\textit{$^\ddagger$Georgia Institute of Technology, Atlanta, GA, USA}\\
*$^\dagger$firstname.lastname@imec.be~
$^\ddagger$cman8@gatech.edu, tushar@ece.gatech.edu
}
}
\maketitle

\begin{abstract}
    \input{text/0_abstract}
\end{abstract}


\input{text/1_introduction}
\input{text/2_background}
\input{text/3_methodology}
\input{text/4_1_experiments}

\input{text/4_case_studies}
\input{text/5_discussion}
\input{text/6_conclusion}

\section*{Acknowledgment}
\noindent {\small This work is funded by the Advanced Research + Invention Agency~(ARIA).}

\input{output.bbl}

\clearpage
\appendix
\input{text/7_artifacts}

\end{document}

%% file: text/0_abstract.tex
Predicting the performance of large-scale distributed machine learning (ML) workloads across multiple accelerator architectures remains a central challenge in ML system design. Existing GPU and TPU focused simulators are typically architecture-specific, while distributed training simulators rely on workload-specific analytical models or costly post-execution traces, limiting portability and cross-platform comparison. This work evaluates whether MLIR’s StableHLO dialect can serve as a unified workload representation for cross-architecture and cross-fidelity performance modeling of distributed ML workloads.
The study establishes a StableHLO-based simulation methodology that maps a single workload representation onto multiple performance models, spanning analytical, profiling-based, and simulator-driven predictors. Using this methodology, workloads are evaluated across GPUs and TPUs without requiring access to scaled-out physical systems, enabling systematic comparison across modeling fidelities. An empirical evaluation covering distributed GEMM kernels, ResNet, and large language model training workloads demonstrates that StableHLO preserves relative performance trends across architectures and fidelities, while exposing accuracy trade-offs and simulator limitations. Across evaluated scenarios, prediction errors remain within practical bounds for early-stage design exploration, and the methodology reveals fidelity-dependent limitations in existing GPU simulators.
These results indicate that StableHLO provides a viable foundation for unified, distributed ML performance modeling across accelerator architectures and simulators, supporting reusable evaluation workflows and cross-validation throughout the ML system design process.

%% file: text/1_introduction.tex
\section{Introduction}
\label{sec:introduction}

The ever-growing compute demands of machine learning (ML) models~\cite{epoch2024trainingcompute}, particularly for large-scale distributed training, have driven rapid advances across both hardware and software. Compute, memory, and network capabilities continue to scale, while the software stack, model architectures, and algorithmic techniques evolve at an unprecedented pace~\cite{yun2025new}. 

\begin{figure}[t]
    \centering
    \includegraphics[width=1\linewidth]{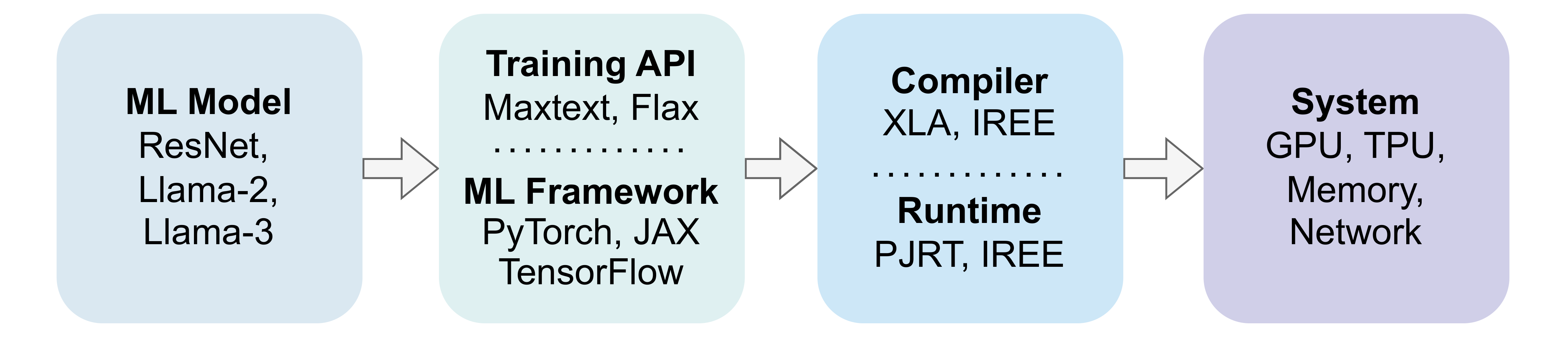}
    \caption{Simplified ML stack overview: from model specification through frameworks, compilation, runtime, and system layers.}
    \label{fig:ml-stack}
\end{figure}

Modern ML systems improve through concurrent innovations across the ML software–hardware stack, as illustrated in Fig.~\ref{fig:ml-stack}. Advances in model architectures~\cite{shazeer2017moe, gu2024mamba, li2025survey}, kernel and compiler optimizations~\cite{Dao2022flashattention,tillet2019triton,chetlur2014cudnn,khan2019miopen,li2024onednn,xla,ansel2024pytorch2}, and hardware platforms~\cite{nvidia2024blackwell,amd2023cdna3,google2025ironwood} interact in complex ways, making performance optimization an inherently multi-dimensional and interdependent problem. In distributed training, these interactions extend beyond a single device to include collective communication, parallelization strategies, and network effects. As a consequence of this co-evolution, understanding and optimizing performance has become increasingly difficult through empirical evaluation alone. Exhaustively benchmarking across model variants, compiler configurations, and hardware platforms is often prohibitively expensive or infeasible, particularly when target systems are unavailable or still under design. Simulation and performance modeling therefore play a critical role in exploring design trade-offs, enabling early-stage evaluation, and providing insight into system behavior that cannot be easily isolated on physical hardware.

In practice, distributed ML training system evaluation relies on a collection of simulation tools operating at different fidelities and targeting different accelerator architectures. Fast analytical models are typically used for early-stage architectural exploration, while higher-fidelity simulators or profiling-based approaches are employed to study detailed performance characteristics on specific hardware platforms. In parallel, GPU and TPU focused simulators have evolved largely independently, each assuming distinct software stacks, workload abstractions, and modeling interfaces. While effective in isolation, such fragmentation is limiting. In particular, workloads, which are the main input, are not easily reproducible or interoperable across distributed ML simulators without substantial re-engineering effort. As workloads transition between fidelities, simulators, or target platforms, they are often reimplemented or approximated, hindering cross-correlation and validation, and making it difficult to attribute observed performance differences to true architectural effects rather than inconsistencies in workload representation or modeling assumptions.

\begin{figure*}[t]
    \centering
    \includegraphics[width=\textwidth]{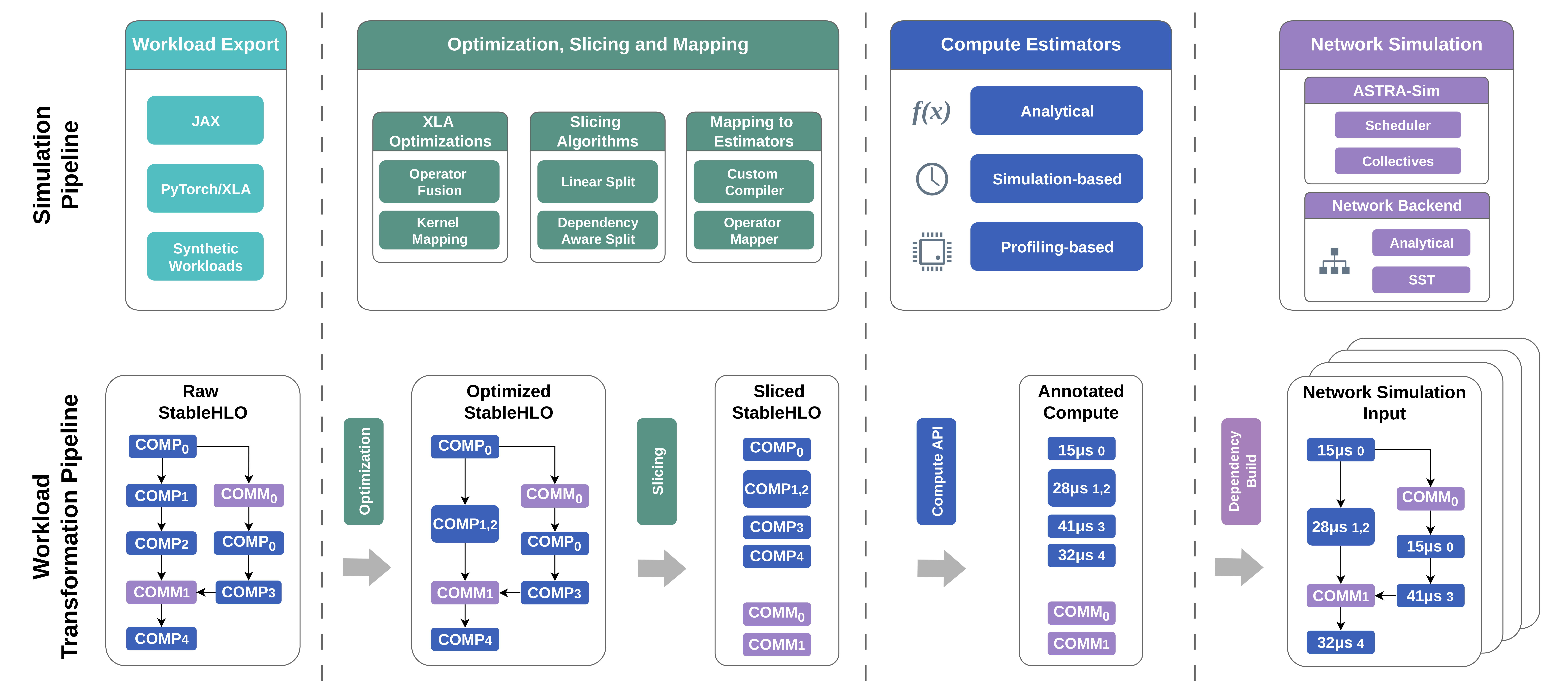}
    \caption{Overview of the StableHLO-based evaluation methodology for distributed ML performance modeling. Workloads are represented in StableHLO, subjected to compiler optimizations and slicing, and mapped through a common compute API to multiple compute backends, including analytical, simulation-based, and profiling-based estimators. The resulting latencies are added to a trace and input to a network simulation backend modeling distributed execution. This workflow enables reuse of a single workload representation across simulation fidelities, accelerator architectures, and modeling tools, supporting cross-correlation and validation of performance results.
    }
    \label{fig:overview_diagram}
\end{figure*}

To address this fragmentation, this work evaluates whether MLIR’s StableHLO~\cite{stablehlo-spec} dialect can serve as a common workload representation for distributed ML performance modeling across accelerator architectures, simulation fidelities, and modeling tools. The evaluation is conducted by establishing a unified simulation methodology, illustrated in Fig.~\ref{fig:overview_diagram}, that uses StableHLO as the workload abstraction and maps it onto multiple simulators of different fidelity levels. This study examines whether representing workloads once in StableHLO and applying the methodology across compute models—ranging from fast analytical estimators to higher-fidelity simulators—can reduce reimplementation effort, enable cross-correlation and validation across fidelities, and support fair comparison across heterogeneous accelerator architectures such as GPUs and TPUs. The feasibility and limitations of this approach are evaluated through case studies spanning workloads, architectures, and performance modeling techniques. This paper makes the following contributions:

\begin{itemize}
    \item Establishes a \emph{StableHLO-based distributed ML simulation methodology} spanning the ML software–hardware stack.
    \item Enables \emph{cross-architecture and cross-fidelity performance modeling}, covering analytical, simulation-based, and profiling-based approaches, using a unified StableHLO representation.
    \item Presents an \emph{empirical evaluation of StableHLO as a unified workload representation} across distributed workloads, accelerator architectures (GPUs and TPUs), and simulator fidelities.
\end{itemize}

This paper is organized as follows. Section \ref{sec:background_succinct} reviews the challenges of existing workload representations and motivates the need for a unified approach. Section \ref{sec:methodology} presents the proposed StableHLO-based methodology and outlines how it interfaces with analytical, profiling-based, and simulation-based predictors. Section \ref{sec:experiments_setup} describes the experimental setup and Section \ref{sec:case_studies} evaluates the approach through case studies on GPUs, TPUs, and multi-node distributed configurations. Section \ref{sec:discussion} discusses limitations, open challenges, and opportunities for future extensions, including integration with the detailed GPU simulators Accel-Sim~\cite{khairy2020accel}. Finally, Section \ref{sec:conclusion} concludes the paper.

\begin{table}[t]
    \caption{Trade-offs between three classes of distributed DNN training simulators: {\em analytical, profiling-based }and {\em simulation-based.}}
\label{tab:sim_type}
{\footnotesize
\renewcommand{\arraystretch}{1.3}%
\setlength{\tabcolsep}{0.5em} 

\begin{tabular}{>{\raggedleft\arraybackslash}p{2cm}p{2cm}p{2cm}p{2cm}}
\toprule
\textbf{Aspect} & \textbf{Analytical} & \textbf{Profiling-based} & \textbf{Simulation-based} \\
\midrule

\textsc{Workload Type} 
  & \raggedright Configuration parameters of high level ML model
  & \raggedright Post-execution traces
  & {\raggedright Executable IR or instruction traces}\\
 \textsc{Workload Flexibility} 
 & \low 
 & \high 
 & \medium\\ \hdashline
 \textsc{Hardware Flexibility}
  & \high 
  & \low 
  & \low  \\
\textsc{Configuration Complexity}
  & \low
  & \medium
  & \high \\
 \textsc{Target Hardware Required}
  & \bad 
  & \good 
  & \bad  \\ \hdashline
\textsc{Fidelity}
& \low 
& \high 
& \high \\
\textsc{Simulation Rate} 
  & \high 
  & \medium
  & \low  \\
 \textsc{Accuracy}
  & \low
  & \high
  & \high \\
 \textsc{Validation Difficulty}
  & \high 
  & \low 
  & \medium \\

\textsc{Bottleneck visibility} & 
\low &
\medium &
\high \\ \hdashline
\textsc{Notes} & 
\raggedright $\oplus$~Fast exploration\\$\ominus$Target-specific compiler opts not considered &
{\raggedright $\oplus$~Captures compiler effects \\
$\ominus$~Framework and hardware coupled \\
$\ominus$ No novel \\architectures supported} &
{\raggedright $\oplus$~Detailed bottleneck visibility\\
$\ominus$ Hard to maintain}\\ 
\hdashline
\textsc{Simulators}
& ~\cite{amped,calculon,optimus,lu2023distsim,ardalani2024deepflow} 
& ~\cite{won2023astrasim2,wang2025simai,gui2025accelerating,bang2024vtrain,shen2025atlahs,proteus,zhu2020daydream,feng2024echo}
& ~\cite{llmcompass} \\
\midrule
\textbf{This work} & \multicolumn{3}{p{6cm}}{
StableHLO driven unified methodology enabling analytical, profiling-based, and simulation-based performance prediction.} \\
\bottomrule

\end{tabular}
}
\end{table}

%% file: text/2_background.tex
\section{Background}
\label{sec:background_succinct}

\noindent Performance modeling for ML system design has led to the development of three broad classes of performance predictors: \emph{analytical}, \emph{profiling-based}, and \emph{simulation-based}. Analytical models (e.g.,~\cite{amped,calculon,optimus,lu2023distsim,ardalani2024deepflow}) operate on coarse-grained abstractions and provide fast, low-cost estimates suitable for early-stage exploration. Profiling-based predictors (e.g.,~\cite{won2023astrasim2,wang2025simai,gui2025accelerating,bang2024vtrain,shen2025atlahs,proteus,zhu2020daydream,feng2024echo}) derive workloads from compiler intermediate representations (IRs) or execution traces, capturing realistic behavior but requiring access to target software stacks or hardware. Simulation-based tools (e.g.,~\cite{llmcompass}) offer higher fidelity and enable evaluation of hypothetical architectures, but incur higher modeling and execution cost. Table~\ref{tab:sim_type} summarizes the trade-offs across these classes.

In practice, these predictor classes are realized through largely independent simulator ecosystems, resulting in fragmented evaluation workflows. A contributing factor to this fragmentation is the workload representation assumed by each simulator, which shapes how compiler transformations, computation, and communication are modeled. As a result, workloads are frequently reimplemented or approximated when moving between analytical, profiling-based, and simulation-based tools, complicating cross-fidelity validation and obscuring whether observed performance differences stem from architectural effects or from inconsistencies in workload representation.

These observations suggest that differences in workload representation play a central role in shaping interoperability, reuse, and comparability across predictor classes. To study whether a single representation can support cross-fidelity and cross-architecture performance modeling, we first articulate the properties such a representation should satisfy. Table~\ref{tab:ir_comparison} summarizes these properties across existing representations, and the following criteria describe the requirements considered in this work.

\begin{table*}[ht]
\footnotesize
\centering
\caption{Comparison of workload representations, their properties, and their simulation capabilities. }
\label{tab:ir_comparison}
\renewcommand{\arraystretch}{1.2}
\setlength{\tabcolsep}{2pt}
\begin{tabular}{rcc>{\centering}p{1.6cm}>{\centering}p{1.5cm}>{\centering}p{1.5cm}cl}
\toprule
\textbf{Representation} & \textbf{Expressivity} & \textbf{Portability} & \textbf{Ahead of Time} & \textbf{Train / Infer} & \textbf{Framework agnostic} & \textbf{Stability} & \textbf{Used in Simulators} \\
\midrule
\papername{\textbf{Configuration-based}} & \low & \bad & \good & \good/\good & \good & \bad &\cite{amped,calculon,optimus,lu2023distsim,ardalani2024deepflow} \\
\hdashline
\papername{\textbf{Post-execution traces:}}\\
\papername{Chakra~\cite{chakra}} & \medium & \bad & \bad & \good/\good & \good & \bad & \cite{won2023astrasim2,llmservingsim}, \textbf{This work} \\ 
\papername{GOAL~\cite{hoefler2009groupgoal}} & \medium & \bad & \bad & \good/\good & \good & \bad & \cite{shen2025atlahs} \\ 
\hdashline
\papername{\textbf{ML Operator-level IRs:}} \\
\papername{DistIR~\cite{distir}} & \high & \bad & \good & \good/\good & \good & \bad & ~\cite{distir} \\ 
\papername{TorchFX~\cite{reed2022torchfx}} & \high & \good & \good $\dagger$ & \good/\good & \bad & \bad & \cite{lee2025forecasting,feng2024echo} \\ 
\papername{ONNX~\cite{onnxbai2019}} & \medium & \good & \good & \bad/\good & \good & \good & \cite{ham2024onnxim,mei2020zigzag} \\ 
\papername{XLA HLO~\cite{openxla_hlo}} & \high & \good & \good & \good/\good & \good & \bad & \nota \\ 
\textbf{StableHLO~\cite{stablehlo-spec}} & \high & \good & \good & \good/\good & \good & \good & \textbf{This work} \\

\bottomrule
\end{tabular}

\raggedright
\vspace{0.2cm}
{ 
\footnotesize  $\dagger$ TorchFX graphs may be obtained via symbolic tracing, but are commonly produced at run time by TorchDynamo's bytecode analysis~\cite{ansel2024pytorch2}. \\
$\oplus$ Some specific backends implement this but not part of the core specifications of the representation. }
\end{table*}

\subsubsection{Expressive}

The representation should be able to express a broad range of workloads executable by common machine learning frameworks. For distributed workloads, it must explicitly represent communication operations, such as collective primitives, which are absent from some formats (e.g., ONNX). To support accurate performance modeling, the representation should capture essential operator semantics without introducing unnecessary complexity. In contrast, high-level configuration-based descriptions are easy to simulate but typically target narrow workload classes and omit performance-critical details.

\subsubsection{Interoperable}
The representation should provide a pathway for integration with analytical, profiling-based, and simulation-based predictors, enabling consistent cross-fidelity evaluation using a single source of truth. Fragmented IR ecosystems force each simulator to maintain custom workload pipelines, which impedes cross-validation and systematic comparison.

\subsubsection{Portable}
The representation should be portable across diverse hardware and runtime environments. It should enable execution through compilation or runtime interpretation on multiple systems for ground-truth validation, ensuring that relevant compiler and runtime optimizations are visible to the simulation.

\subsubsection{Ahead-of-Time}
The workload description should be available without requiring execution on the target system. Trace-based representations, such as Chakra execution traces~\cite{chakra} or \texttt{Dynamo}-generated \texttt{torch.fx} graphs, are therefore less suitable, as they depend on prior execution and restrict modeling flexibility.

\subsubsection{Capture a Full Training Step}
The representation must include the forward pass, backward pass, and optimizer update to reflect the complete training workload. Formats primarily designed for inference, such as ONNX, limit their applicability for end-to-end training simulation.

\subsubsection{Framework-Agnostic}
The representation should be generatable from multiple ML frameworks, improving usability and adoption. Framework-specific solutions such as \texttt{torch.fx} lack this flexibility.

\subsubsection{Stable}
Rapid IR evolution, including the introduction of new operators and execution patterns, introduces forward-compatibility risks. Stability helps ensure reproducibility and long-term viability of workload descriptions.

%% file: text/3_methodology.tex
\section{Methodology}
\label{sec:methodology}

\input{text/3_1_stablehlo_subsec}

\noindent \textbf{(d) Network Simulation:}

Once the compute regions’ latencies are modeled, we proceed to full system simulation. Distributed ML Network simulators, such as ASTRA-sim~\cite{won2023astrasim2} and ATLAHS~\cite{shen2025atlahs}, accept a trace-graph representation in which vertices denote compute or communication collectives and edges denote data dependencies. These simulators model the workload scheduler and network behavior to estimate end-to-end performance for distributed workloads.

In this study, we use ASTRA-sim’s analytical backend~\cite{won2023astrasim2} for simple network configurations and to enable detailed network modeling we extend ASTRA-sim with an SST~\cite{sst2011acm} network backend. The performance-annotated StableHLO graph is mapped to the Chakra~\cite{chakra} format, a version-controlled ML trace format adopted by MLCommons. Chakra traces are then ingested by ASTRA-sim: each compute region is mapped to a \texttt{COMP} node, communication operators are mapped to \texttt{COMM} nodes, communication sizes are inferred from tensor types, and communication semantics are derived from the corresponding StableHLO collective operator.

%% file: text/3_1_stablehlo_subsec.tex
\noindent In this section, we describe the StableHLO-based simulation methodology evaluated in this work and outline how it enables cross-architecture and cross-fidelity performance modeling.

\subsection{StableHLO as a Unified Workload Representation}

StableHLO~\cite{stablehlo-spec,stablehlo-github} provides a fixed, versioned operator set with precise semantics. When extended with sharding annotations, such as \texttt{mhlo.sharding} or operations from OpenXLA’s \texttt{sdy} dialect~\cite{openxla_sdy_dialect}, it captures computation, data partitioning, and collective communication semantics required for distributed training workloads. Implemented within MLIR~\cite{lattner2021mlir}, StableHLO exposes programmatic interfaces and transformation utilities that integrate with existing compiler passes. This abstraction level is well suited for performance modeling: operators expose tensor shapes, layouts, and communication patterns without incurring kernel- or instruction-level trace overhead. A fragment of StableHLO syntax is shown in Fig.~\ref{lst:IR-example}.

\begin{figure}[h]
\centering
\begin{lstlisting}[basicstyle=\footnotesize\ttfamily]
%0 = stablehlo.dot_general %a, %b {
  lhs_contracting_dimensions = [0],
  rhs_contracting_dimensions = [0]
} : (tensor<4x6xf32>, tensor<4x3xf32>)
  -> tensor<6x3xf32>

%1 = stablehlo.transpose %0 {
  permutation = [1, 0]
} : (tensor<6x3xf32>)
  -> tensor<3x6xf32>

%2 = stablehlo.all_reduce %1 (
  ^bb0(%x: f32, %y: f32):
    %s = stablehlo.add %x, %y : f32
    stablehlo.return %s : f32
) : (tensor<3x6xf32>)
  -> tensor<3x6xf32>
  ...
\end{lstlisting}
\caption{Fragment of StableHLO-MLIR illustrating a matrix multiplication, followed by a transpose and an all\_reduce collective.}
\label{lst:IR-example}
\end{figure}
\begin{figure}[ht]
    \centering
    \includegraphics[width=0.6\linewidth]{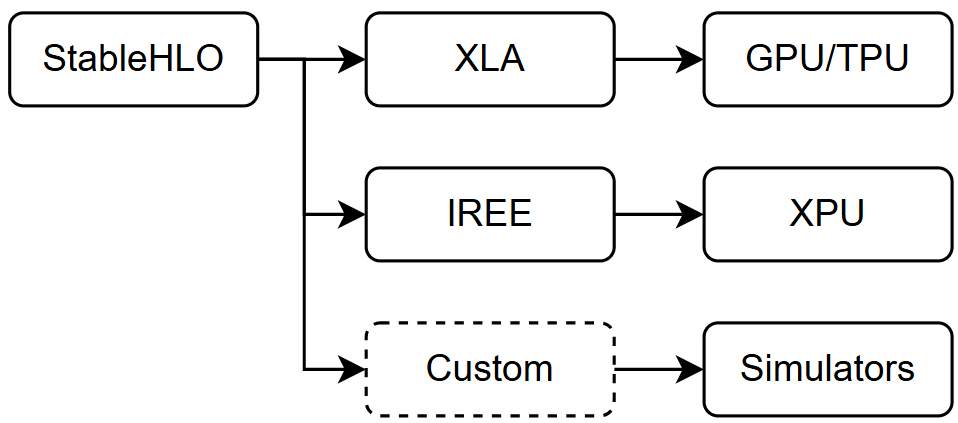}
    \caption{Compiler pathways enabled by StableHLO-MLIR. StableHLO serves as a common input to multiple compiler toolchains, including XLA, IREE, and custom flows.}
    \label{fig:compiler_options}
\end{figure}

This representation can be consumed by analytical models, simulation tools, or compiled for profiling via toolchains such as XLA or IREE (Fig.~\ref{fig:compiler_options}), enabling a unified evaluation pipeline across simulator classes.

\subsection{Unified Performance Evaluation Methodology}
Building on StableHLO as the common workload representation, we now describe the performance evaluation methodology used in this study (illustrated in Fig.~\ref{fig:overview_diagram}), which proceeds in four stages:
\begin{enumerate}[(a)]
\item \emph{Workload Export,}
\item \emph{Workload Optimization, Slicing and Mapping,}
\item \emph{Compute Estimation,}
\item \emph{Network Simulation}.
\end{enumerate}

\noindent \textbf{(a) Workload Export:}
Workloads are exported as distributed StableHLO with sharding annotations and collectives preserved from reference implementation libraries (e.g., MaxText\footnote{MaxText~\cite{maxtext} provides ready-to-use distributed training implementations of large language models with configurable model size and parallelism, however any frontend that lowers to StableHLO can be used}). Workloads may also be constructed directly in StableHLO to enable targeted studies via parameterized microbenchmarks.

\noindent \textbf{(b) Optimization, Slicing and Mapping:}
This stage performs three transformations on the StableHLO workload: (i) compiler optimizations; (ii) workload slicing and communication extraction; and (iii) mapping to a compute estimator.

For analytical and simulation-based simulators, a raw StableHLO export does not represent real compiled execution. Typically, compilation for a real hardware backend entails multiple stages of optimization and code generation. Since StableHLO is compatible with multiple compiler toolchains (e.g., XLA, IREE), optional compiler passes can be applied to the workload to approximate hardware-specific compilation effects and obtain a more representative input for performance modeling.

For the slicing stage we distinguish compute and communication operators because they are typically consumed by different simulators: compute operators (e.g., matrix multiplication, convolution, elementwise operations) are evaluated on compute hardware simulators, while communication operators (e.g., all\_reduce, all\_gather) usually drive the network simulation stage. This separation enables independent modeling of accelerator execution and system-level communication, allowing compute performance to be evaluated without requiring full system simulation.

To enable this, we identify communication primitives and separate them from computation using two slicing algorithms.
The first is a \emph{linear split} that partitions the program into alternating communication and compute regions following dependency order. All consecutive compute operations between communication primitives are grouped into a single compute region, resulting in larger regions with minimal analysis overhead. This approach is well suited for profiling-based estimators, as it enables a broader compilation scope and exposes more compiler optimization opportunities. The second is a \emph{dependency-aware split} that operates at the individual operator level, capturing exact data dependencies. This typically produces smaller compute regions, which are more suitable for analytical or simulation-based modeling and may expose additional compute–communication overlap. 

Finally, each compute region is mapped to a compute latency estimator to obtain an execution latency.

\noindent \textbf{(c) Compute Estimation:}
Three pathways are supported: (i)~analytical modeling, where operator-level metrics are aggregated (e.g., a roofline model over the region); (ii)~simulation-based estimation, where regions are evaluated using a compute simulator; and (iii)~profiling-based estimation, where regions are compiled and executed on target hardware to obtain measured runtimes.

To make this process consistent across estimators, we expose a minimal \emph{Compute API} that each estimator implements:
\begin{itemize}
  \item \texttt{get\_run\_time\_estimate()}: returns the estimated latency for a compute region. Internally, this may compile and execute the region or evaluate an analytical model such as roofline over its operators.
  \item \texttt{get\_compile\_args()} (optional): supplies compiler flags and configuration (e.g., target, passes) when using compilation and profiling.
  \item \texttt{get\_exec\_args()} (optional): supplies runtime flags (e.g., number of runs, etc.) for execution or simulator runs.
\end{itemize}
This shared interface also enables mixing multiple compute latency estimators within the same workload—useful when some operators are unsupported on a given estimator (e.g., systolic-array simulators limited to matrix multiplications)—while preserving a single point of control for latency collection. By decoupling workload IR from the estimator, the methodology supports evaluation across fidelity levels: analytical models, detailed simulators, and real devices (GPU, CPU, TPU).

To reduce evaluation cost, we exploit the repetition in modern ML models (e.g., stacked transformer blocks) by caching latency results per compute region. The caching key is the tuple \emph{(target hardware \(H\) $\times$ compilation toolchain \(C\) $\times$ compute region \(R\))}, since changing any of \(H\), \(C\), or \(R\) can affect latency. This eliminates redundant simulations and lowers runtime, especially for profiling and simulation-based estimation. On profiling runs, we observe this caching mechanism to result in an 89.7\% average reduction in evaluation time on Llama-3 and 26.8\% on ResNet workloads. 

At the end of this stage, the workload is annotated with latency values for each compute region, making it ready for system-level simulation.

%% file: text/4_1_experiments.tex
\section{Experimental Setup}\label{sec:experiments_setup}

\subsection{Workload Setup}
\subsubsection{Workload Selection} 
We evaluate three representative distributed training workloads: data-parallel ResNet training, fully sharded data-parallel Llama-3 training, and large-scale data-parallel Llama-2 training following ATLAHS~\cite{shen2025atlahs} scale-out configurations as a reference. Workloads are exported to StableHLO from JAX~\cite{jax} during just-in-time compilation of the training step. ResNet workloads are generated using reference examples from the \texttt{flax} library~\cite{flax2020github}, while Llama workloads are generated using the MaxText library~\cite{maxtext}. The workloads and configurations used in this study are summarized in Table~\ref{tab:workloads}.

\label{sec:export}

\begin{table}[t]
\centering
\footnotesize
\caption{Overview of workloads and configurations, considering DP and FSDP parallelization strategies.}
\label{tab:workloads}
\setlength{\tabcolsep}{1pt}
\begin{tabular}{lrcrrrr}
\toprule
\textbf{Workload} & \textbf{\#Parameters} & \textbf{Datatype} & \textbf{Nodes} & \textbf{Batch size/device} & \textbf{Parallelism} \\
\midrule
ResNet 18-200    & 11–65M & FP16  & 4         &  256 &   DP \\
Llama-3.1        & 0.1–3B & BF16  & 4         &    1 & FSDP \\
Llama-2          &     7B & BF16  & 16--128   & 2--1 &   DP \\
\bottomrule     
\end{tabular}
\end{table}

\begin{table}[t]
\centering
\footnotesize
\caption{GPU system and analytical roofline parameters used in the evaluation.}
\label{tab:gpu_system_and_roofline}
\setlength{\tabcolsep}{5pt}
\begin{tabular}{lccc}
\toprule
\textbf{GPU Model} & \textbf{Peak Compute} & \textbf{Memory BW} & \textbf{NVLink BW} \\
 & \textit{TFLOP/s} & \textit{TB/s} & \textit{GB/s} \\
\midrule
A100 (40GB SXM)   & 312   & 1.94 & 100  \\
H100 (80GB SXM)   & 1979  & 3.35 & 150  \\
H200 (141GB SXM)  & 1979  & 4.80 & 150  \\
B200 (180GB HGX)  & 4500  & 7.70 & 300  \\
\midrule
\textbf{GPU count}     & \multicolumn{3}{c}{4 GPUs} \\
\textbf{Topology}       & \multicolumn{3}{c}{All-to-all NVLink} \\
\midrule

\textbf{GPU count}   & \multicolumn{3}{c}{16/128 GPUs} \\
\textbf{Topology, Link}    & \multicolumn{3}{c}{Intranode: all-to-all (NVLink)} \\
{}                   & \multicolumn{3}{c}{Internode: dragonfly (Slingshot)} \\

\bottomrule
\end{tabular}
\end{table}

\begin{figure}[t]

\begin{subfigure}[t]{0.4\columnwidth}
    \centering
    \caption{}
    \includegraphics[width=\columnwidth]{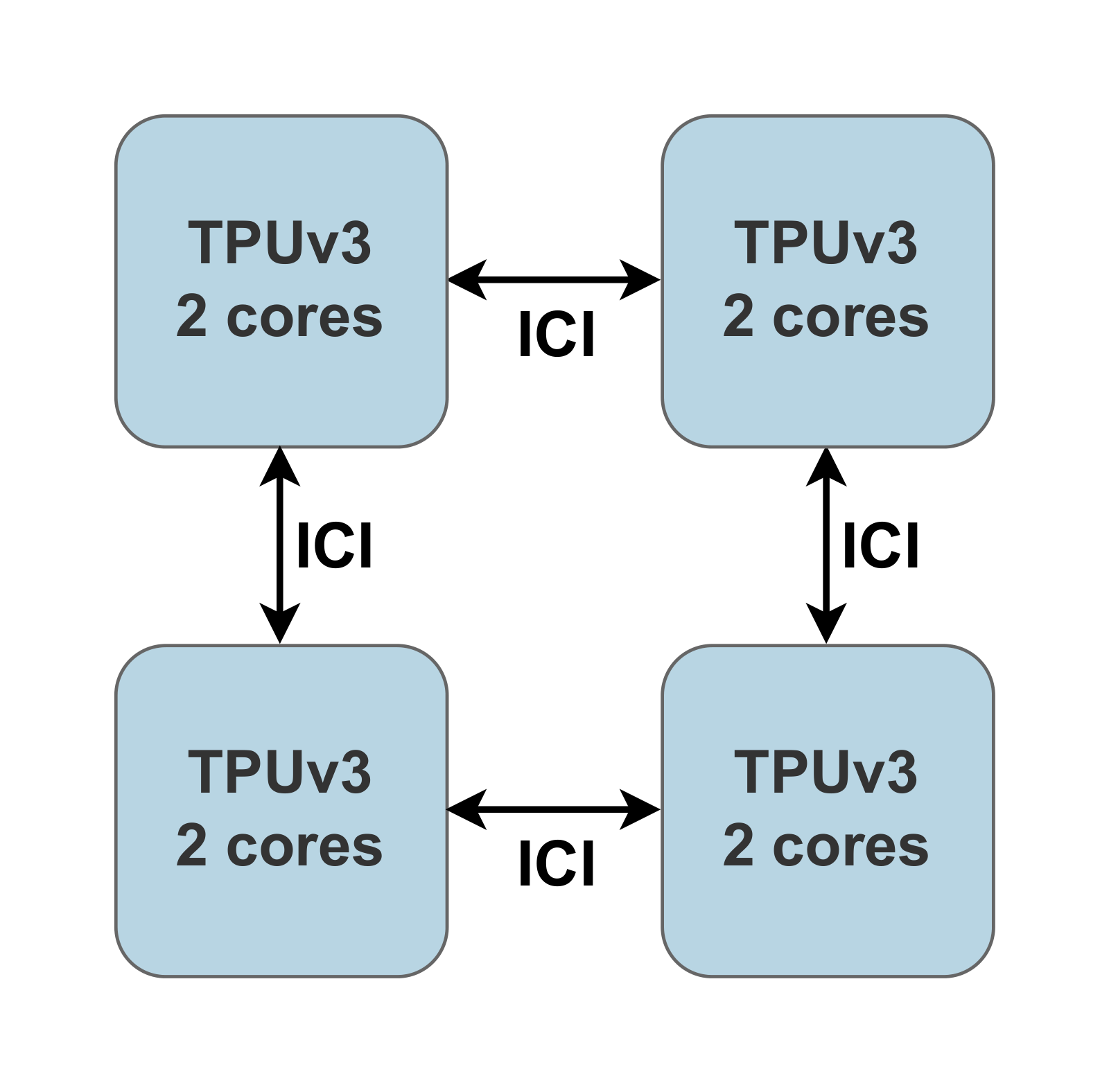}
\end{subfigure}
\begin{subfigure}[t]{0.4\columnwidth}
    \centering
    \footnotesize
    \caption{}
    \vspace{8pt}
    \begin{tabular}{rl}\toprule
        \textbf{Parameter} & \textbf{Value}\\ \midrule
        Cores        & 8 \\
        Systolic Array/Core & 2 \\
        Peak Compute$^\dagger$    & 63.3 TFLOP/s \\
        Peak Memory BW$^\dagger$  & 429.2 GB/s \\
        Clock Rate      & 940 MHz \\
        System Topology & Mesh \\
        ICI BW/link     & 656 Gb/s \\ \bottomrule
    \end{tabular}
\end{subfigure}

\caption{TPUv3-8 system diagram and key parameters. $^\dagger$ Per-core peak compute and memory bandwidth values are taken from OpenXLA's \texttt{xprof}~\cite{openxla_xprof_2025} tool.}
\label{fig:tpuv3_ref}
\end{figure}

\subsubsection{Optimization pipeline}
We apply selective optimization passes to StableHLO workloads using OpenXLA's \texttt{hlo-opt} tool to capture realistic compilation effects. The optimization pipeline consists of three stages: (1) lowering StableHLO to HLO via \texttt{xla-translate}, (2) optimizing with \texttt{hlo-opt} in deviceless mode, and (3) raising the optimized HLO back to StableHLO. 

All standard GPU passes remain enabled except those related to autotuning and asynchronous collectives. Disabling autotuning limits optimizations to those that do not require hardware feedback. Passes related to asynchronous collectives are disabled because they are incompatible with the system-simulation input construction methodology.

\subsection{Validation Platforms}
\subsubsection{GPU setup} We consider three GPU system configurations with 4, 16, and 128 GPUs for reference executions used to validate the proposed methodology. Table~\ref{tab:gpu_system_and_roofline} summarizes the corresponding system-level parameters.

\subsubsection{TPU setup} We reserve an 8-core TPUv3 pod slice for our reference runs. The system configuration is shown in Fig.~\ref{fig:tpuv3_ref}.

\subsubsection{Software stack} Workloads on both GPUs and TPUs are deployed using the JAX-based implementations described~\ref{sec:export}. We use the timing collected from the deployment of these workloads as our references. Reference measurements for TPUs were measured using the \texttt{xprof}~\cite{openxla_xprof_2025}, and cross-checked against~\cite{lewis2022large}.

\subsection{Simulation Setup}
\subsubsection{Analytical Estimator} The analytical estimator implements a per-operator roofline model~\cite{roofline}, using peak FLOP/s and peak memory-bandwidth parameters to estimate each operator’s runtime and selecting the dominant bottleneck. Fused regions are modeled as a single compute region, with memory traffic accounted only at the region boundaries while preserving the full compute cost of all constituent operations. This formulation allows optimized StableHLO inputs to be consumed directly and yields substantially improved accuracy relative to the raw exports.

\subsubsection{Profiling Estimator} For the GPU backend, the profiling estimator executes each compute region on actual hardware using \texttt{hlo\_runner\_main} from the OpenXLA toolchain to obtain measured runtimes. Profiling latencies are averaged over five runs to reduce measurement noise. Due to the closed-source nature of the TPU runtime, we do not use a profiling-based estimator for TPUs.

\subsubsection{Simulation-based Estimators} Two systolic-array simulators are selected for the TPU evaluation, ONNXim~\cite{ham2024onnxim} and COCOSSim~\cite{choudhary2025cocossim}, and integrated as compute estimators via the Compute API. Because these tools primarily target matrix-multiplication workloads, they are paired with an analytical TPU estimator to model StableHLO operators outside their native support. Both simulators are configured to approximate the architecture of a single TPUv3 core~\cite{Jouppi2020TPUSupercomputer} as closely as their configurability allows, enabling controlled comparison of simulation fidelity and cost. 

\subsubsection{Network Simulator}
For 4-GPU and TPU configurations, collective communication is modeled using ASTRA-sim’s analytical network backend, representing NVLink and the TPU inter-chip interconnect (ICI), respectively. For 16- and 128-GPU systems, we use ASTRA-sim integrated with SST-Merlin~\cite{sst2011acm} as a network simulation backend to enable packet-level network simulation. ASTRA-sim schedules collective operations derived from StableHLO communication operators, while SST-Merlin performs detailed interconnect modeling.

\subsubsection{Evaluation metrics} We report simulated runtime (the predicted execution time), simulation runtime (wall-clock time), and mean absolute percentage error (MAPE) relative to a hardware reference run (ground truth).

%% file: text/4_case_studies.tex
\section{Case Studies}\label{sec:case_studies}
\subsection{4-Node GPU Configuration} 
\begin{figure}[t]
    \includegraphics[width=\linewidth]{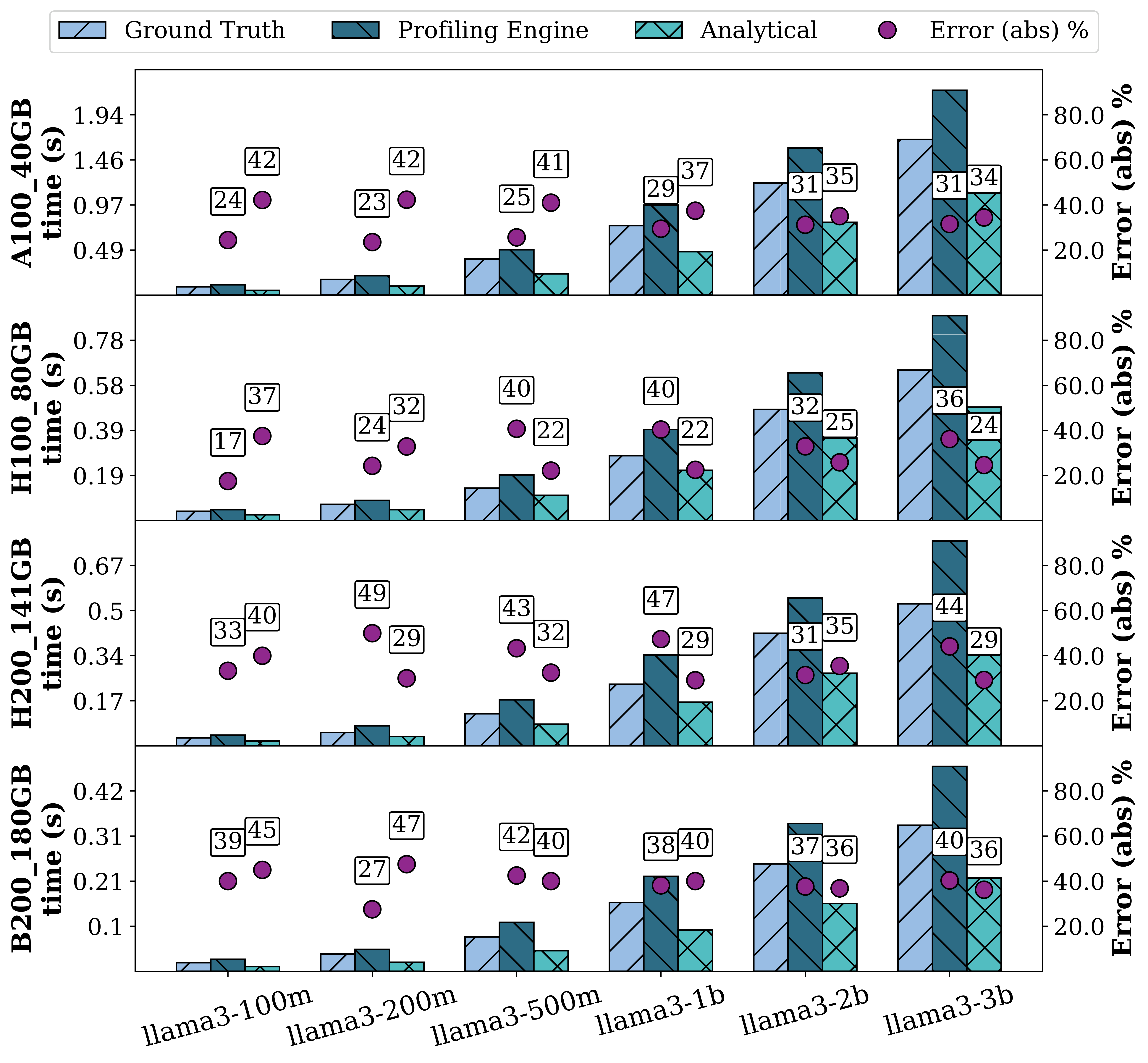}
    \caption{Training-step latency estimates of single training step of \llama variants on a 4-GPU node across three NVIDIA GPU generations. Each color denotes a different performance estimator: analytical models using optimized StableHLO workloads, a profiling-based estimator, and a reference ground-truth measurements. Absolute percentage errors are shown alongside each estimator.}
    \label{fig:gpu_llama3}
\end{figure}

We first evaluate the proposed methodology across multiple GPU generations (A100, H100, H200, and B200), using a common StableHLO representation of a Llama-3 training workload. All workloads are exported in configurations targeting four GPUs. We model a 4-GPU system using ASTRA-sim’s analytical backend, with all-to-all NVLink connectivity between GPUs, representative of the reference system configurations summarized in Table~\ref{tab:gpu_system_and_roofline}. The experimental results are shown in Fig.~\ref{fig:gpu_llama3}.

Across all GPU generations, the analytical estimator achieves approximately 35\% MAPE, while the profiling-based estimator achieves 25--40\% MAPE for the Llama-3 workloads. As expected analytical runs are optimistic whereas profiling runs pessimistic with respect to the ground truth. A primary source of error in the profiling-based estimator is tied to OpenXLA compiler’s inability to apply global, end-to-end optimizations across the sliced compute region boundaries.

To assess the relative performance trends across hardware generations, we compute speedups across GPU generations. Speedup between platforms is defined as $S = T_{\text{prev}}/T_{\text{next}}$, 
and the relative speedup error measures deviation from the reference:
\begin{equation}
\epsilon_S = \frac{S_{\text{Reference}} - S_{\text{Simulated}}}{S_{\text{Reference}}}
\end{equation}
Table~\ref{tab:speedup_error} shows speedup errors averaged across 
Llama-3 workloads. The profiling-based estimator achieves lower speedup 
errors (mean absolute 4.3\%) compared to the analytical model (12.0\%), 
indicating it better captures architectural scaling trends.

\begin{table}[t]
\centering
\caption{Relative speedup error across GPU generations, averaged over 
Llama-3 workloads.}
\label{tab:speedup_error}
\begin{tabular}{lcc}
\toprule
\textbf{Transition} & \textbf{Profiling} & \textbf{Analytical} \\
\midrule
A100 $\rightarrow$ H100  & 3\%    & 15\%  \\
H100 $\rightarrow$ H200  & 7\%    & $-7\%$ \\
H200 $\rightarrow$ B200  & $-3\%$ & 14\%  \\
\midrule
Mean Absolute & 4.3\% & 12.0\% \\
\bottomrule
\end{tabular}
\end{table}

A comparison of simulation runtimes highlights the cost--accuracy tradeoff between the two estimators. Analytical roofline evaluations complete in under 12 seconds for Llama-3 workloads, making them suitable for rapid design-space exploration. In contrast, profiling-based estimation requires several minutes (median $\approx 377$s), due to repeated compilation and on-device execution of compute regions. This gap underscores the value of supporting multiple fidelity levels within a unified methodology: analytical models enable fast estimates, while profiling-based approaches capture detailed kernel-level behavior at increased cost, though their accuracy can be affected by the loss of end-to-end compiler optimizations when workloads are evaluated in isolation.

\begin{figure}[t]
    \centering
    \includegraphics[width=0.85\linewidth]{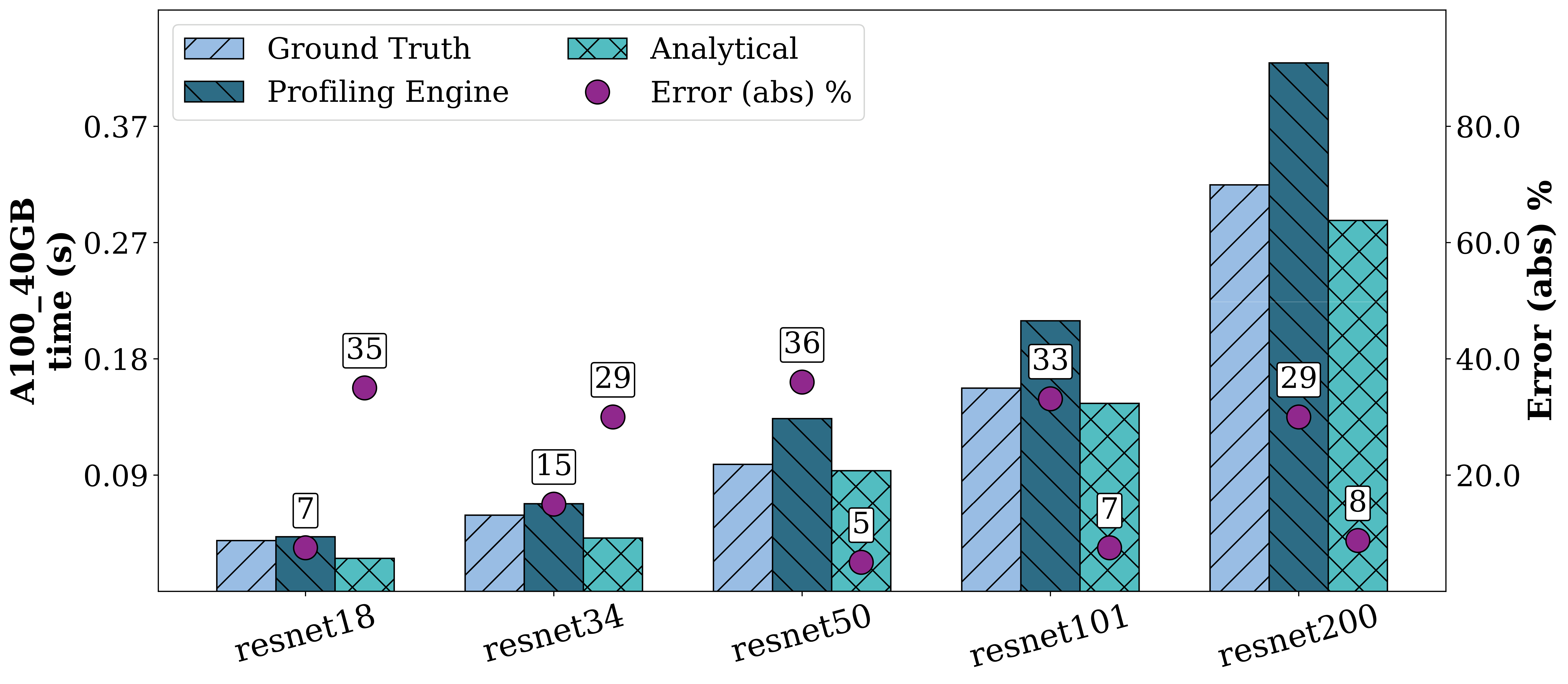}
    \caption{ Training-step latency estimate  for ResNet variants on a 4-GPU A100 system.}
    \label{fig:eval_resnet}
\end{figure}

Fig.~\ref{fig:eval_resnet} shows simulation results for multiple ResNet variants using the same compute estimator configuration on a 4-GPU A100 system. This illustrates that the proposed StableHLO-based methodology generalizes beyond LLM workloads to conventional deep neural network training.

\subsection{Large-scale GPU Configuration (16-128 GPUs)}
To evaluate the applicability of StableHLO as a workload representation at scale, we study large-scale distributed training configurations spanning tens to hundreds of GPUs.

\begin{figure}[t]
    \centering
    \begin{subfigure}{\columnwidth}
        \centering
    \caption{Illustration of the simulated dragonfly topology.}
    \label{fig:dragonfly}
    \includegraphics[width=1.12\columnwidth]{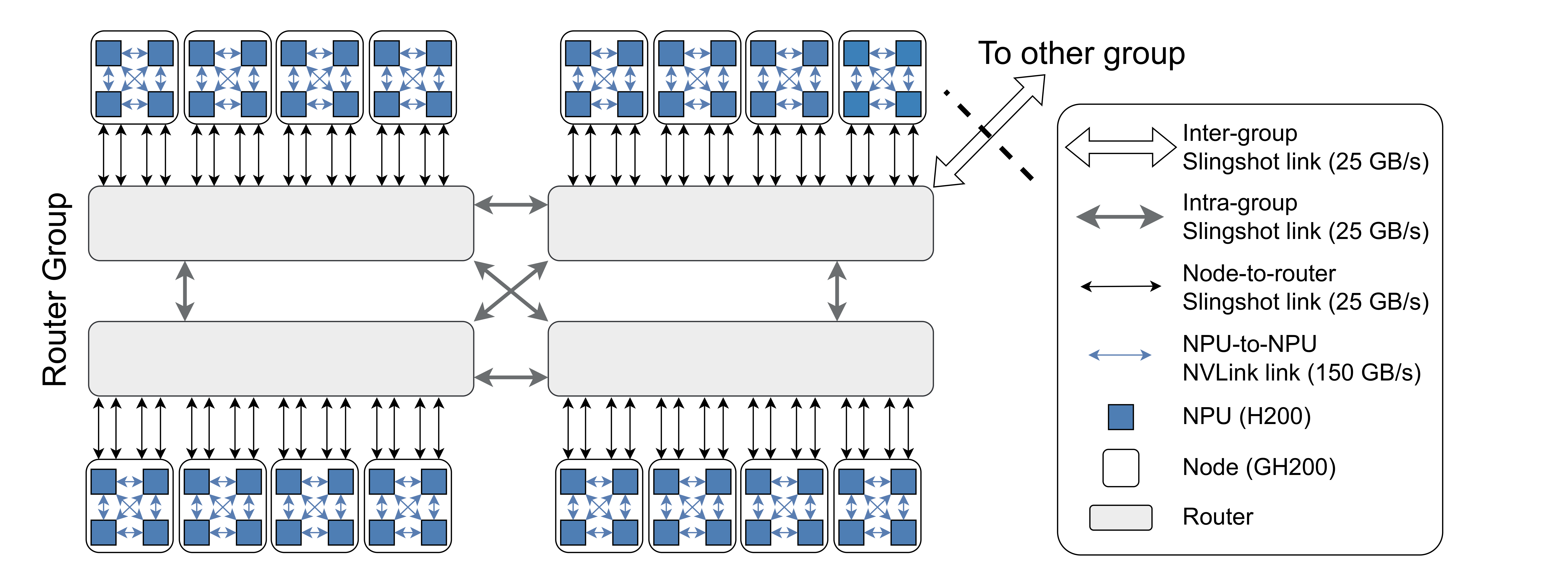}
    \end{subfigure}
\begin{subfigure}{\columnwidth}
\caption{Dragonfly topology parameters}
\label{tab:scaleout_topo}
\footnotesize
    \begin{tabular}{ccccc}
\toprule
\textbf{\#Nodes} & \textbf{GPUs/node} & \textbf{Nodes/router} & \textbf{Routers/group} & \textbf{Groups} \\
32 & 4 & 4 & 4 & 2 \\
4 & 4 & 1 & 2 & 2 \\
\bottomrule
\end{tabular}
\end{subfigure}
\caption{Large-scale GPU system with dragonfly topology}
\label{fig:large_gpu}
\end{figure}

We consider the Llama-2 7B workload using the system configurations reported in ATLAHS~\cite{shen2025atlahs}. Each node consists of four GH200 GPUs connected via an all-to-all NVLink fabric, with nodes arranged in a dragonfly topology~\cite{kim2008technologydragonfly}. The modeled topology parameters are summarized in Fig.~\ref{fig:large_gpu}.

\begin{figure}[t]
    \centering
    \includegraphics[width=0.7\linewidth]{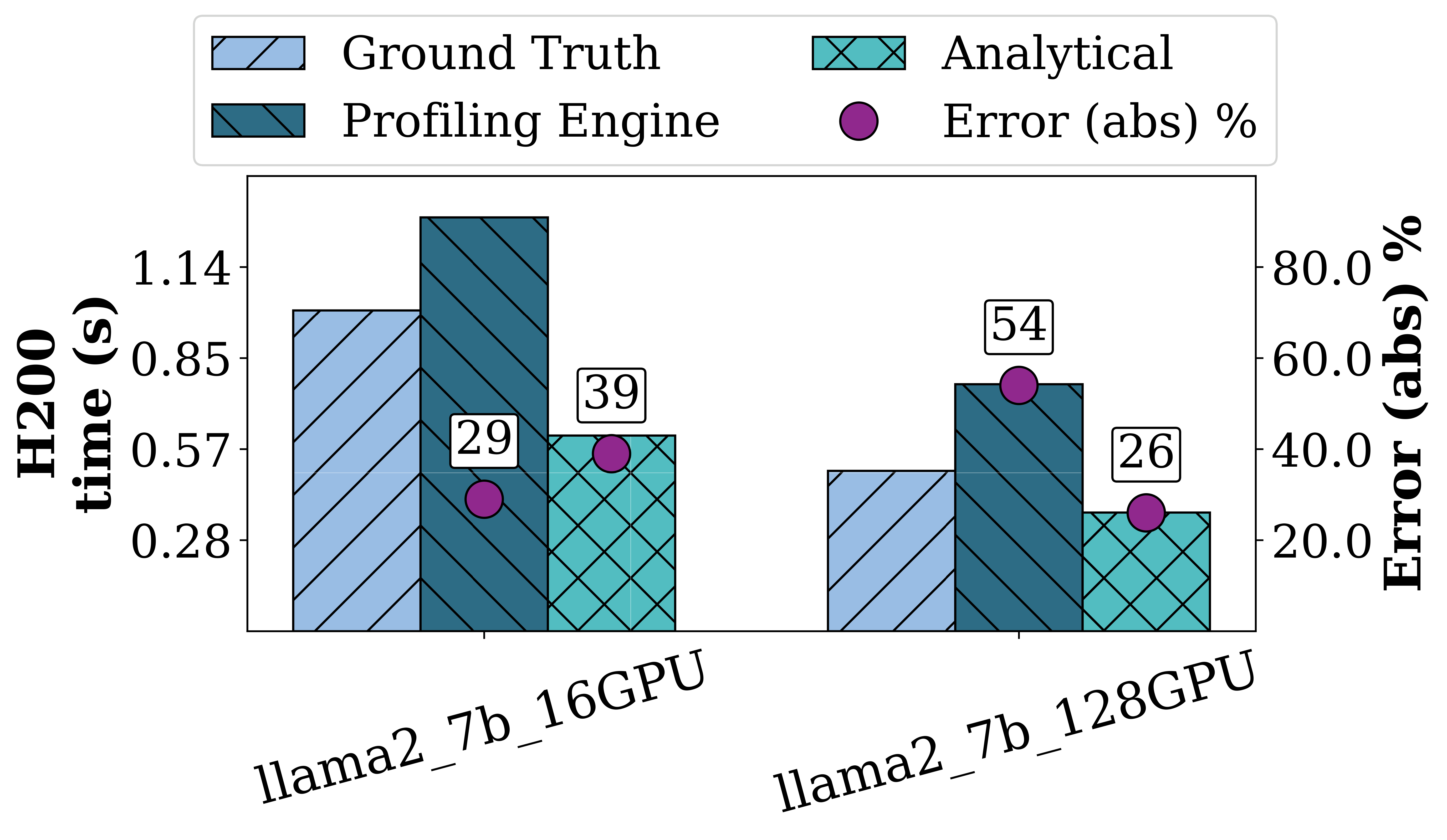}
    \caption{Llama-2 scale-out simulation results using ASTRA-sim with an SST-Merlin~\cite{sst2011acm} network backend for 16- and 128-GPU systems, compared against reference results from~\cite{shen2025atlahs}.}
    \label{fig:llama2-scaleout}
\end{figure}

The scale-out results in Fig.~\ref{fig:llama2-scaleout} show that both compute estimators reproduce the qualitative scaling trends reported in ATLAHS. At 16 GPUs, the analytical estimator exhibits a 39\% error relative to the reference, while the profiling-based estimator achieves a lower deviation of 29\%. At 128 GPUs, the analytical estimator remains comparatively stable with a 26\% error, whereas the profiling-based approach becomes less reliable, yielding a 54\% deviation. This degradation is attributed to reduced compiler optimization opportunities when compute regions are evaluated in isolation across deeper communication hierarchies.

Despite differences in absolute accuracy, both estimators capture the relative increase in communication cost with scale. These results demonstrate that StableHLO enables end-to-end scale-out simulation across large GPU systems while preserving global performance trends of distributed training workloads across modeling fidelities.

Across all evaluated GPU configurations, the analytical roofline estimator consistently produces optimistic latency estimates relative to the hardware reference, while the profiling-based estimator consistently overestimates execution time. This behavior reflects the complementary biases of the two approaches: the roofline model is based on peak performance limits, whereas profiling-based estimation incurs pessimism due to compute-region isolation and the loss of end-to-end compiler optimizations. As a result, reference measurements  fall between the analytical and profiling-based estimates.

\subsection{TPUv3 system evaluation (8 cores)} \label{subsec:tpu}
We next evaluate the methodology on TPUs~\cite{jouppi2017datacenter} to assess whether the StableHLO workloads can also drive a fundamentally different accelerator architecture. This experiment demonstrates how a single StableHLO representation enables consistent evaluation across heterogeneous compute latency estimators.

\begin{figure}
    \centering
    \includegraphics[width=1\linewidth]{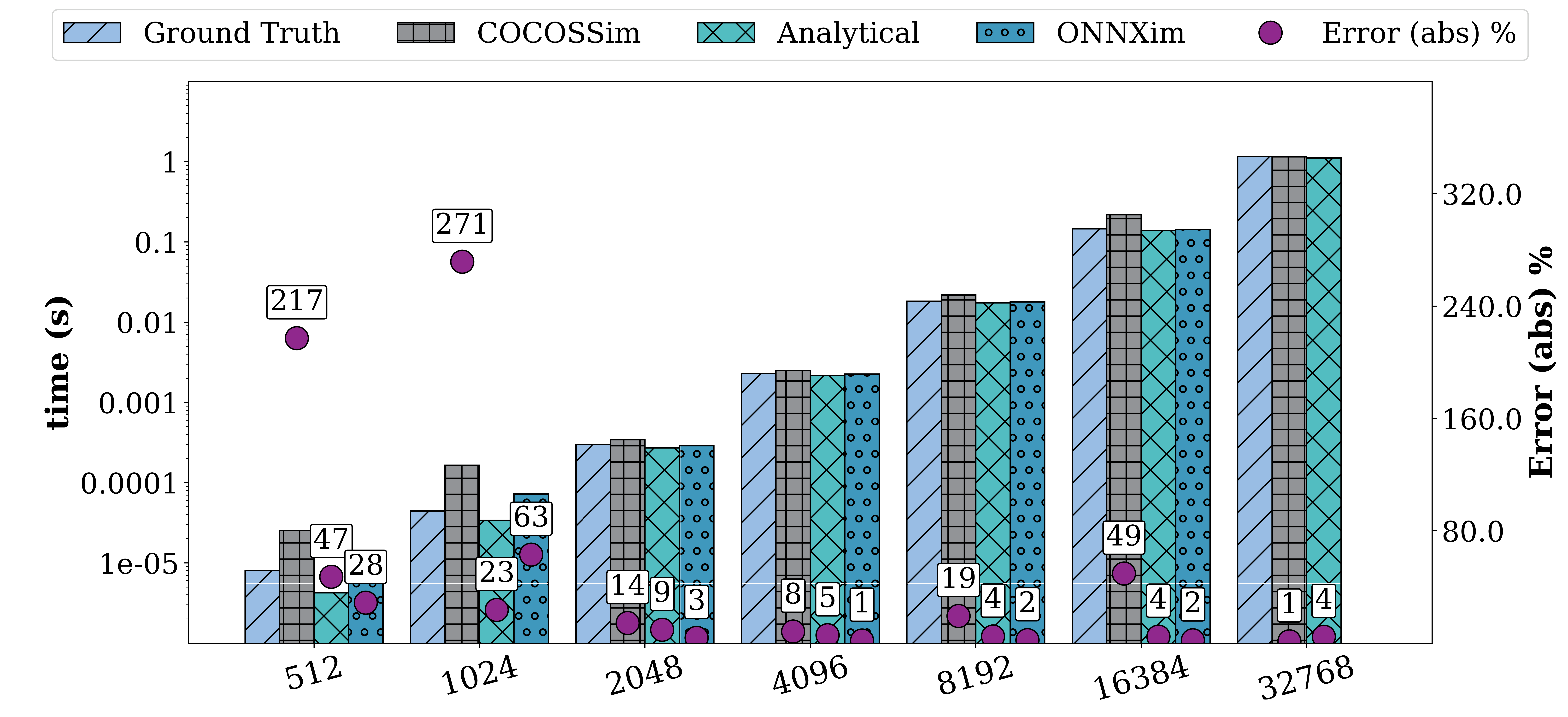}
    {\footnotesize 
\begin{tabular}{lrr}
\toprule
\textbf{Simulator}  &  \textbf{Simulation Time}~(s) & \textbf{MAPE}~(\%) \\  \midrule
ONNXim      & 20724.63  & 2.47 \\
COCOSSim   & 477.72 & 18.29 \\ \bottomrule
\end{tabular}
}
    \caption{Operator-level benchmarking of matrix multiplication~(expressed as StableHLO \texttt{dot\_general} operator) with  square matrix dimensions $M=N=K$, across four systolic-array simulators, each configured to approximate a TPUv3 core with two systolic arrays.}
    \label{fig:tpu_matmul}
\end{figure}

\subsubsection{TPU matrix multiplication study}
We evaluate four systolic-array simulators: COCOSSim~\cite{choudhary2025cocossim}, ONNXim~\cite{ham2024onnxim}, SCALE-Sim~\cite{raj2025scale_sim_v3}, and ZigZag~\cite{zigzag} using matrix-multiplication operation across a range of problem sizes, expressed as {\em dot\_general} StableHLO operator. As shown in Fig.~\ref{fig:tpu_matmul}, ONNXim and COCOSSim most closely track TPUv3 latency trends for large GEMMs, achieving mean absolute percentage errors of approximately 2\% and 18\%, respectively. In contrast, SCALE-Sim and ZigZag exhibit substantial deviations, resulting in large errors even at larger matrix sizes.

Simulation cost varies significantly across tools. ONNXim delivers high accuracy but requires several hours to simulate large matrix multiplications, whereas COCOSSim provides a more favorable accuracy--runtime tradeoff and remains practical for full-model studies. Based on this operator-level analysis, COCOSSim is selected as the primary systolic-array estimator for end-to-end TPU experiments.

\begin{figure}[t]
    \centering
    \includegraphics[width=1\linewidth]{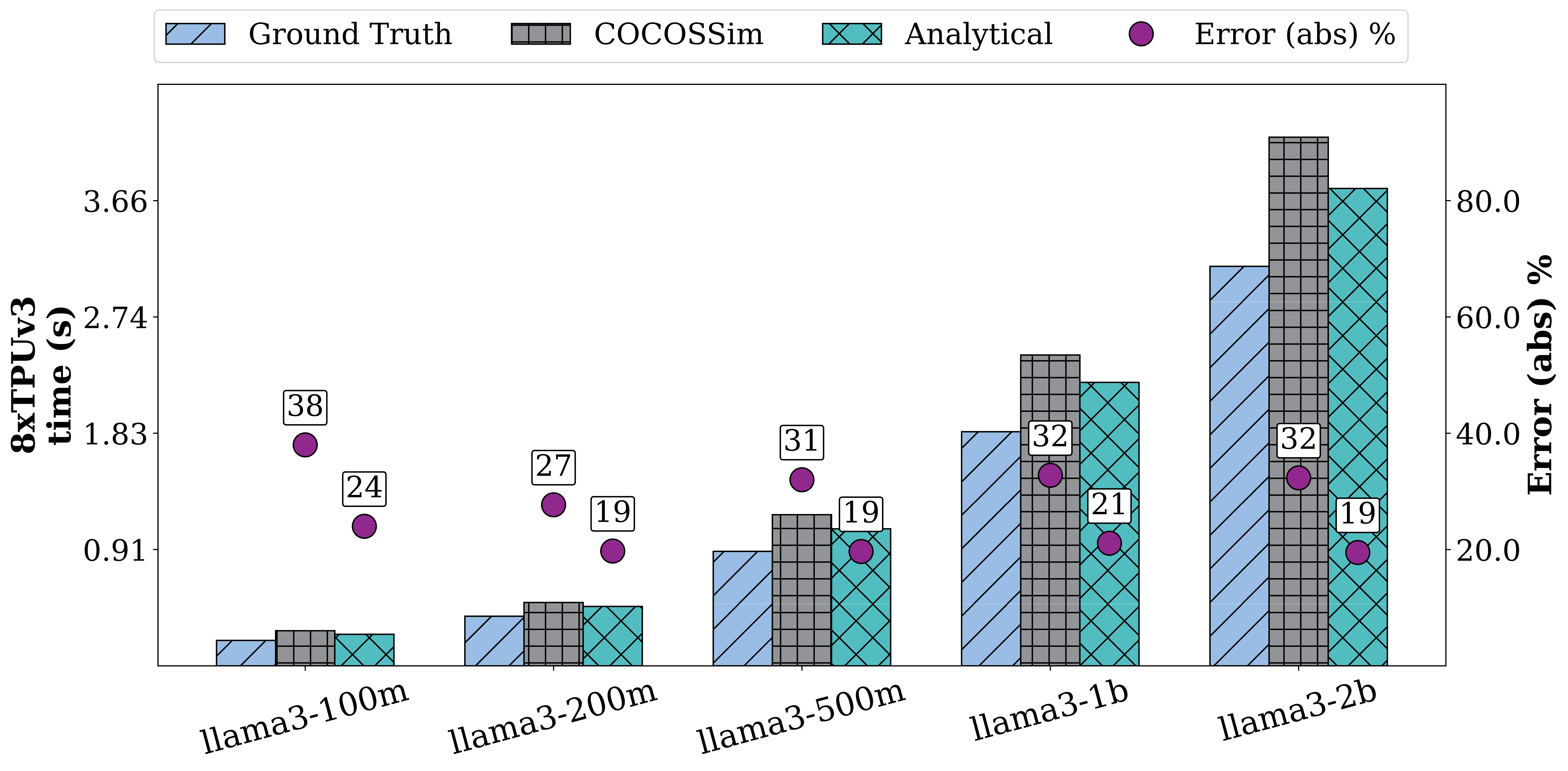}
    \caption{\llama Training Step across different TPU performance estimators. Mean runtime for the 
    analytical and COCOSSim estimator is 6.42s and 826.55s respectively.}
    \label{fig:tpuv3_system}
\end{figure}

\subsubsection{End-to-end evaluation} Building on the operator-level results, we next evaluate full Llama-3 training-step workloads on an 8-core TPUv3 system using both COCOSSim and an analytical TPU estimator.
 
As the XLA compilation toolchain for TPU backends is closed source, we use the optimized StableHLO variant obtained via \texttt{hlo-opt} using the GPU configuration as an input to the analytical estimator. This approximation relies on the assumption that key compiler transformations, particularly fusion and common subgraph simplifications, are largely device-agnostic.

System-level results for Llama-3 training steps are shown in Fig.~\ref{fig:tpuv3_system}. The analytical TPU estimator achieves the closest alignment with reference TPU runtimes, yielding absolute errors of 19--24\% across all model sizes. The COCOSSim backend also captures the overall performance trend but exhibits higher deviations, with errors of 27--38\% for the optimized representation.

Runtime differences between the estimators are substantial: the analytical model completes in a few seconds per workload, whereas COCOSSim requires several minutes depending on model scale. In this case study, the simulation-based estimator does not improve accuracy relative to the analytical model. A primary contributing factor is the lack of visibility into TPU-specific compiler decisions, such as scratchpad memory allocation and scheduling, which likely leads to pessimistic predictions compared to the reference.

This experiment highlights the flexibility of the proposed methodology. The same StableHLO workload executes unmodified across multiple compute estimators, including a detailed systolic-array simulator, enabling the integration of diverse performance models within a unified evaluation framework.

%% file: text/5_discussion.tex
\section{Discussion} \label{sec:discussion}
\begin{figure}[t]
    \centering
    \includegraphics[width=\linewidth]{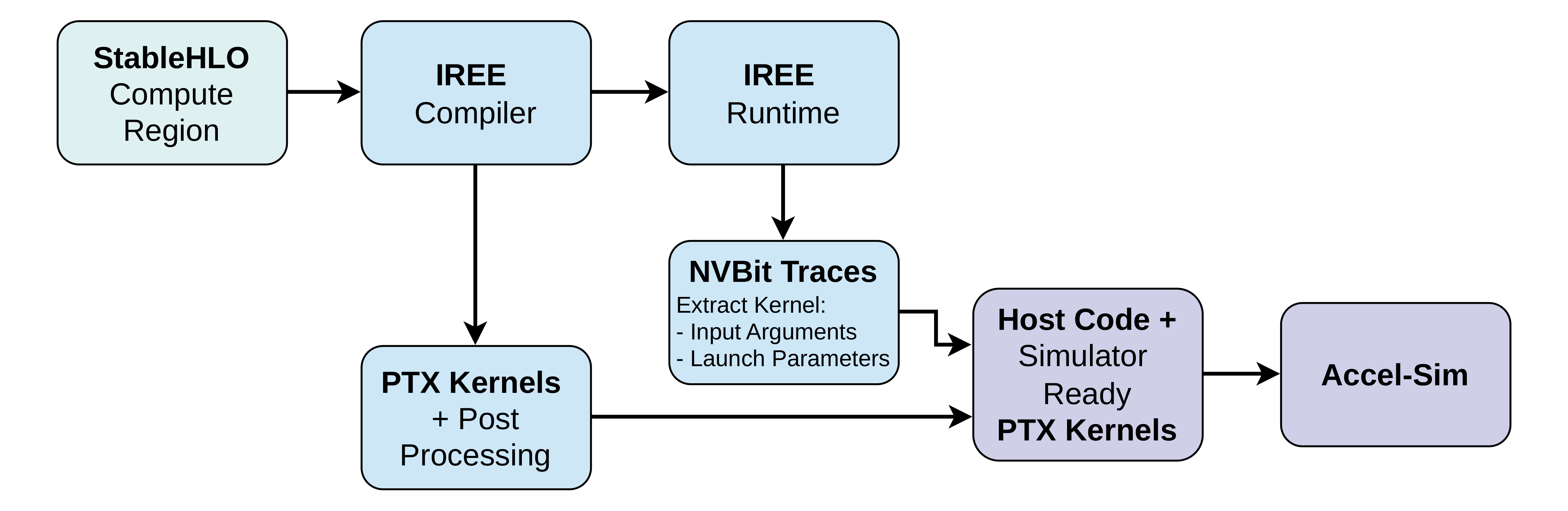}
    \caption{IREE--Accel-Sim integration flow used to evaluate StableHLO workloads with detailed GPU simulation.}
    \label{fig:accelsim-flow}
\end{figure}
\subsection{Evaluating StableHLO for Detailed GPU Simulation}
Detailed architectural simulators such as Accel-Sim~\cite{khairy2020accel} provide instruction-level visibility that is inaccessible on real GPUs, making them valuable for microarchitectural exploration. However, integrating such simulators into modern ML compiler toolchains remains challenging. To evaluate whether StableHLO can serve as a viable input representation for detailed GPU simulation, we developed the workflow shown in Fig.~\ref{fig:accelsim-flow} and integrated it into our methodology via the Compute API.

The workflow translates StableHLO programs into PTX suitable for Accel-Sim simulation. Although XLA is used elsewhere in this work, this flow relies on the IREE~\cite{iree} compiler. OpenXLA frequently dispatches to precompiled SASS kernels (e.g., cuDNN and cuBLAS), which are incompatible with PTX-level tracing. In contrast, IREE can be configured to generate all kernels in PTX. However, IREE does not expose kernel launch parameters or runtime inputs. To recover this information, we intercept kernel launches using an NVBit-based instrumentation tool while executing the program against the NVIDIA stub driver, allowing IREE to proceed as if running on real hardware.

Due to the high cost of Accel-Sim simulation, we evaluate this flow using a small transformer model with approximately 4M parameters, exported from Flax examples~\cite{flax2020github}. Even at this scale, significant practical challenges arise. Of the 42 kernels generated for this workload, 38 required manual PTX modifications due to Accel-Sim parsing limitations, unsupported instructions, or failures related to dynamic shared-memory allocation. Ultimately, only 20 kernels could be simulated successfully.

Simulation runtimes were also substantial. The median simulation time for a single kernel exceeded 2.1 hours, with several kernels requiring more than 15 hours to complete. These results indicate a fundamental mismatch between detailed GPU simulators and the rapidly evolving compiler stacks and kernel generation strategies used by modern ML frameworks. While Accel-Sim remains a powerful tool for targeted microarchitectural studies, {\em using Accel-Sim for end-to-end ML training workloads, even at modest scale, remains impractical}.

Taken together, these findings motivate the need for intermediate-fidelity GPU performance models that balance accuracy, compiler compatibility, and scalability. StableHLO enables such models by providing a portable, compiler-compatible workload representation.

\subsection{Limitations of StableHLO for Simulation}

While StableHLO provides a unified and compiler-compatible workload representation, using it as the foundation for distributed ML performance simulation exposes several limitations. These limitations stem largely from the fact that StableHLO was designed as a high-level interchange format for XLA, rather than as a complete or self-contained specification for detailed performance modeling.

Although StableHLO enables interaction with XLA’s optimization pipeline, many performance-critical transformations such as: operator fusion, layout decisions, and sharding propagation, are performed internally on XLA’s native HLO representation. These transformations may introduce operations or semantics that are not representable in StableHLO. As a result, exporting a post-optimization StableHLO workload requires selectively disabling compiler passes via \texttt{hlo-opt} to avoid constructs that cannot be raised back into StableHLO. This process depends on undocumented interactions between passes and can lead to the loss of important optimization effects, limiting the fidelity of post-export representations and making it difficult to trace just-in-time compilation behavior.

In addition, practical StableHLO exports often rely on auxiliary MLIR dialects, such as SDY or MHLO, to express semantics that are not fully captured in StableHLO itself, including sharding annotations and certain operators. Support for these dialects is inconsistent outside the OpenXLA ecosystem; for example, some compiler toolchains have deprecated MHLO support entirely. Consequently, StableHLO cannot function as a fully standalone workload representation and instead requires a tightly coupled compilation environment to remain usable and accurate.

These limitations indicate that StableHLO is best viewed as a compiler-compatible workload abstraction rather than a complete simulation IR. While it enables reuse across modeling fidelities and architectures, achieving high-fidelity simulation still requires careful control of compiler pipelines and an understanding of where information is lost across representation boundaries.

\subsection{JAX vs. PyTorch}
While PyTorch remains the dominant ML framework and recent work~\cite{feng2024echo, lee2025forecasting, pytorchsim},  
has explored simulation using PyTorch-based representations, several factors 
motivated the choice of StableHLO and JAX for this study. First, the \texttt{torch.compile} 
API is less mature than the StableHLO/XLA compilation stack for compile-driven 
parallelism and static graph transformations. \texttt{torch.compile} cannot 
guarantee a complete ahead-of-time graph for all workloads due to graph breaks 
caused by untraceable control flow, which is particularly limiting for complex 
training workloads such as mixture-of-experts models. Second, while ahead-of-time 
compilation is supported for PyTorch inference via \texttt{torch.export}, 
full AoT compilation of training workloads---including backward and optimizer 
update phases---is an active area of development. In contrast, StableHLO 
naturally represents a complete training step. Third, StableHLO's portability 
across compilation backends (XLA, IREE, custom flows) provides flexibility 
critical for cross-architecture evaluation. Finally, if the PyTorch frontend 
is indispensable, StableHLO can still be emitted from PyTorch via 
\texttt{torch\_xla}, the compilation path used to execute PyTorch programs 
on TPUs.

%% file: text/6_conclusion.tex
\section{Conclusion}
\label{sec:conclusion}

\noindent This work evaluated whether MLIR’s StableHLO dialect can serve as a unified workload representation for distributed ML performance modeling across a variety of
modeling tools, simulation fidelities and 
compute architectures. The results demonstrate that StableHLO enables a single workload description to drive analytical, profiling-based, and simulation-based estimators for GPUs and TPUs, supporting consistent cross-fidelity evaluation without requiring repeated workload reimplementation.

Across GPU and TPU case studies, the methodology captures global performance trends and scaling behavior while exposing clear accuracy–cost tradeoffs between estimator classes. Analytical models provide fast, optimistic estimates suitable for early-stage exploration, while profiling- and simulation-based estimators offer higher fidelity at increased computational cost and with biases introduced by workload partitioning and limited compiler visibility. Importantly, reference measurements consistently fall between analytical and profiling-based estimates, highlighting the complementary nature of these approaches when used within a unified framework.

The evaluation also exposes inherent limitations of StableHLO as a simulation input. While it enables portable workload reuse and access to mature compiler optimizations, StableHLO does not encode backend-specific scheduling, memory placement, or kernel-level execution details. These gaps limit the effectiveness of detailed architectural simulators and explain why higher-fidelity models do not consistently outperform analytical estimators in this study.

Overall, StableHLO enables reusable, cross-architecture performance evaluation workflows that span multiple simulator classes and fidelity levels. By decoupling workload representation from performance estimation, the methodology supports systematic comparison and validation of distributed ML systems.

%% file: output.bbl

%% file: text/7_artifacts.tex
\section*{Artifact Appendix}

\subsection{Abstract}
This artifact appendix describes how to access and install the simulation framework introduced in Figure~\ref{fig:overview_diagram}. It also describes how to reproduce the experiments in Section~\ref{sec:case_studies} (Figures~\ref{fig:gpu_llama3},~\ref{fig:eval_resnet}, and~\ref{fig:tpuv3_system}). All input configuration files and datasets are included in the artifact.

\subsection{Artifact check-list (meta-information)}
{\small
\begin{itemize}
  \item {\bf Program:} hespas - simulation framework for distributed ML training
  \item {\bf Model:} ResNet variants, Llama-3 (100M--3B), Llama-2 7B
  \item {\bf Data set:} StableHLO-MLIR of ResNet-18/34/50/101/200, Llama-3 (100M--3B), Llama-2 (7B)
  \item {\bf Run-time environment:} Linux-based distribution
  \item {\bf Execution:} Automated through bash scripts and configuration files
  \item {\bf Metrics:} Training step time
  \item {\bf Output:} Log files, CSV summaries, graphs (PDF, PNG, SVG)
  \item {\bf Experiments:} Figures~\ref{fig:gpu_llama3},~\ref{fig:eval_resnet}, and~\ref{fig:tpuv3_system}
  \item {\bf How much disk space required (approximately)?:} 5~GB
  \item {\bf How much time is needed to prepare workflow?:} Approx. 5~min
  \item {\bf How much time is needed to complete experiments?:} Approx. 25~min
  \item {\bf Publicly available:} Yes
  \item {\bf Code licenses:} MIT
\end{itemize}
}

\subsection{Description}

\subsubsection{How to access} 
The artifact for results reproduction is available at \href{https://doi.org/10.5281/zenodo.18874691}{https://doi.org/10.5281/zenodo.18874691}.
The source code is planned to be open-sourced on GitHub at \href{https://github.com/imec-int/hespas}{https://github.com/imec-int/hespas}.

\subsubsection{Hardware dependencies} 
The core simulation framework can be run on any Linux-based system. To reproduce the profiling-based estimator results, matching GPU and TPU hardware is required as specified in Table~\ref{tab:gpu_system_and_roofline} and Figure~\ref{fig:tpuv3_ref}.

\subsubsection{Software dependencies}
The core simulation framework is written in Python 3.10+. For network simulation, it uses ASTRA-sim and therefore depends on the following two repositories:
\begin{itemize}
    \item ASTRA-sim~\cite{won2023astrasim2} for network modeling,
    \item Chakra~\cite{chakra} as the trace format input for ASTRA-sim.
\end{itemize}
    
The following systolic array estimators are utilized as compute estimators in the TPU case study (Section~\ref{subsec:tpu}):
\begin{itemize}
    \item COCOSSim~\cite{choudhary2025cocossim},
    \item ONNXim~\cite{ham2024onnxim}.
\end{itemize}

Note that only COCOSSim is used in the system-level experiments.

The framework has been tested on Ubuntu 22.04.

All paths in this appendix are relative to the root directory of the distribution.

\subsubsection{Data sets}
The simulation framework takes two types of inputs:
\begin{itemize}
    \item System configuration files,
    \item StableHLO ML workloads.
\end{itemize}

All datasets are included in the release package.

\subsection{Installation}
The framework is released as a self-contained package. Upon extraction, the root directory contains a single bash script that installs all required dependencies and executes the experiments. Users may be prompted to install additional system dependencies (e.g., \texttt{cmake} and \texttt{protobuf}). To install and run all experiments, execute the following command:

\begin{verbatim}
    ./run_all.sh
\end{verbatim}

\subsection{Experiment workflow}
To reproduce all artifact-supported paper results, execute: \texttt{./run\_all.sh}. 

This is the only required command. It will automatically:
\begin{itemize}
    \item Build all required components (HeSPaS, ASTRA-sim, COCOSSim).
    \item Run paper experiments.
    \item Generate CSV summaries and final graphs.
\end{itemize}

After completion, all outputs are placed in \texttt{results/}: \texttt{*.csv} for raw experiment data and \texttt{*.png}, \texttt{*.svg}, \texttt{*.pdf} for graphs.

The main run script will automatically simulate the StableHLO input files using the compute estimators and produce a Chakra trace for each configuration, which is then fed into ASTRA-sim for system simulation. The final training step time is accumulated into a CSV file and a graphing script is used to produce the final plots.

The experiment scripts containing the inputs and setup are found under \texttt{repos/hespas/experiments/paper/}.

\textbf{Note on profiling-based estimator:} The profiling-based estimator results found in Figures~\ref{fig:gpu_llama3} and~\ref{fig:eval_resnet} require a GPU to profile the StableHLO workloads. For convenience, we provide Chakra traces pre-annotated with the profiling-based estimator results, e.g., \texttt{repos/hespas/experiments/paper/llama3/\allowbreak profiling/\allowbreak workloads/\allowbreak A100/\allowbreak llama3-100m/}. This allows reproduction of results without GPU access.

A Docker instance with JAX installation is provided for users who wish to re-run the profiling-based estimator with their own GPU hardware. Configuration details can be found in \texttt{README\_docker.md}.

\subsection{Evaluation and expected results}

The provided run flow reproduces the core paper experiments: Figures~\ref{fig:gpu_llama3},~\ref{fig:eval_resnet}, and~\ref{fig:tpuv3_system}.
Figure~\ref{fig:llama2-scaleout} is not part of the reproducible artifact as it relies on a closed-source SST-Merlin-based network simulator. The TPU microarchitectural meta-study in Figure~\ref{fig:tpu_matmul} is excluded from the artifact evaluation scope.

\subsection{Experiment customization}
There are several options for experiment customization.
\begin{itemize}
    \item \textbf{Inputs:} the simplest way to change experiments is to modify the system configuration files by changing the hardware or network topology parameters. Examples of system configuration files can be found in \texttt{repos/hespas/configs}.
    \item \textbf{Compute estimators:} users may add additional compute estimators or choose to modify existing ones. The Compute Estimator API is implemented in \texttt{repos/hespas/src/\allowbreak hespas/estimator}, where users may also find examples of existing estimators.
    \item \textbf{Network Simulators:} The simulation framework uses ASTRA-sim as its network backend. Refer to the ASTRA-sim~\cite{won2023astrasim2} documentation to add new or extend existing network simulators via the ASTRA-sim Network API.
\end{itemize}